\documentclass[a4paper,14pt]{article}

\usepackage{multicol}
\usepackage{amsthm}

\usepackage[dvips]{graphicx,color}
\usepackage{amssymb}
\usepackage{amsmath}

\begin{document}

\fontsize{14pt}{16.5pt}\selectfont

\begin{center}
\bf{Universal topological representation of geometric patterns}
\end{center}
\fontsize{12pt}{11pt}\selectfont
\begin{center}
Shousuke Ohmori$^{1*}$, Yoshihiro Yamazaki$^{1,2}$, Tomoyuki Yamamoto$^{1,2}$, Akihiko Kitada$^{2}$\\ 
\end{center}

\noindent
$^1$\it{Faculty of Science and Engineering, Waseda~University, 3-4-1 Okubo, Shinjuku-ku, Tokyo 169-8555, Japan}\\
$^2$\it{Institute of Condensed-Matter Science, Comprehensive Resaerch Organization, Waseda University,
3-4-1 Okubo, Shinjuku-ku, Tokyo 169-8555, Japan}\\

\noindent
*corresponding author: 42261timemachine@ruri.waseda.jp\\
~~\\
\rm
\fontsize{11pt}{14pt}\selectfont\noindent

\noindent
{\bf Abstract}\\
We discuss here geometric structures of condensed matters by means of a fundamental topological method. 
Any geometric pattern can be universally represented by a decomposition space of a topological space consisting of the infinite product space of $0$ and $1$, in which a partition with a specific topological structure determines a character of each geometric structure. \\

\section{Introduction}
\label{sec:1}

Geometrical structures of condensed matters constituted of atoms or molecules found in nature have been widely studied in disordered systems\cite{Ziman,Cusack}. 
In these studies, the mathematical methods that are independent of the group theory because of an absence of characters of long-range order as to periodicity and symmetry in disorder systems have been developed.  
In particular, various methods using topological concept are proposed as a useful tool for analyzing geometrical configuration directly of atoms or molecules in condensed matters\cite{Monastyrsky,Yoshida}. 
For example, a method of persistent homology which is one of topological date analysis is effective in investigating classification of geometric structures formed by amorphous materials\cite{Hirata,Hiraoka}. 
Note that the method is mathematically based on a technique of algebraic topology.

In several topological methods, we have been successfully studied the mathematical structures of condensed matters by using a fundamental topological approach, that is, a point set topology\cite{Kitada2005,Kitada2007,Kitada2016}. 
By means of this topological method an universal geometric structure of condensed matters which is independent of each detailed nature of structure of them has been investigated qualitatively. 
For instance, we discussed a hierarchic structure of self-similar structures in materials and verified the existence of such hierarchic structure emerging universally into any dendritic structure in which each self-similarity is characterized by Cantor-set.   
Note in these studies that, we have employed a somewhat indirect method of observation of the structures of condensed matters. 
That is, the geometrical structures are expressed indirectly through the mathematical observations of the formation of a set of equivalence classes (such a concept was proposed by Fern$\acute{a}$ndez\cite{Fernandez} in statistical physics).
Here the collection of subsets of a topological space relative to an equivalence classes is called a decomposition space in a point set topology\cite{Nadler,decomposition}.
Note that diffraction analysis is based on the idea of the equivalence class\cite{Cassels}. 
In fact, the group of the lattice plane is gathered as a concept of equivalence class, and then the geometrical structure in a real space can be determined using diffraction patterns in a reciprocal space. 
The geometric structures of condensed matters are observed here mainly from the viewpoint of this equivalence class, namely, decomposition space, which is a similar idea to that of diffraction analysis.

Recently, the authors proposed the mathematical sufficient condition for an issue in material science and geology that a polycrystal can be filled with an arbitrarily finite number of self-similar crystals\cite{Kitada2016}.
According to the issue, if a geometric structure of polycrystal is characterized by a topological space $X$ which has a specific topological structure, namely, $X$ is a 0-dim\cite{zero-dim}, perfect, compact Hausdorff-space, then the mathematical procedure of construction of an arbitrarily finite number of self-similar crystals filling with the polycrystal is ensured. 
In fact, assuming the element of the crystal to be the map $x:\Lambda \to \{0,1\}$, let $X$ be a Cantor cube $(X,\tau)=(\{0,1\}^{\Lambda},\tau_0^{\Lambda}),$ Card $\Lambda \geq \aleph_0$ where a Cantor cube $(\{0,1\}^\Lambda ,\tau_0^\Lambda )$ is the 
$\Lambda -$product space of $(\{0,1\},\tau_0)$ for a set $\Lambda$ and $\tau_0$ is a discrete topology for 
$\{0,1\}$. 
Note that a Cantor cube $(\{0,1\}^{\Lambda},\tau_0^{\Lambda})$ with Card $\Lambda \geq \aleph_0$ is a 0-dim, perfect, compact Hausdorff-space. 
Then, it is confirmed that there exists a partition $\{X_1,\dots,X_n\}$ of $X$ such that each $X_i$ is also a 0-dim, perfect, compact Hausdorff-space and a decomposition space $\mathcal{D}_{X_i}$ of $X_i$ is self-similar.
Therefore, regarding each subspace $X_i$ as a crystal, we obtained a polycrystal composed of crystals each of which is characterized by its self-similar decomposition space. 
We also verified that in this discussion the self-similarity of each crystal can be replaced with a compact substance in materials such as a dendrite, and then we obtained a polycrystal filled with an arbitrarily finite number of crystals with a dendritic structure. 
In the above mathematical procedure, geometric structures of compact substances such as self-similar and dendritic can be represented by a collection of subspaces of $X$ i.e., a decomposition space of $X_i$.
Nevertheless, the practical form of a decomposition space is not yet clear.

In this paper, we discuss geometric structures of condensed matters from the viewpoint of a point set topology. 
Especially, we demonstrate a practical representation of a decomposition space of a Cantor cube $(X,\tau)=(\{0,1\}^{\Lambda},\tau_0^{\Lambda})$ corresponding to each of geometric models with compactness. 
In the next section, we show basic concepts that any compact metric space is represented by a decomposition space of a Cantor cube $X$, the partition of $X$ with 0-dim, perfect, and compact characterizing the practical form of the decomposition space.
Also, we provide a decomposition space homeomorphic to a closed interval $[0,1]$ in real line as a simple case.
In Section 3, we discuss some geometric models with compactness, e.g., graphic, dendritic, clusterized structures and apply the obtained results to the previous paper stated above.
A conclusion is given in Section 4.
\\
\section{Basic concepts}
\label{sec:2}

Any compact metric space is, in principle, obtained homeomorphically by a decomposition space of 0-dim, perfect, compact Hausdorff-space\cite{AH-theorem}. 
In this section, we will show a practical construction of a decomposition space homeomorphic to a compact metric space as a general procedure.

Let $X$ denote a Cantor cube $(\{0,1\}^\Lambda ,\tau_0^\Lambda )$ with Card $\Lambda \geq \aleph _0$, and let $Y$ be a compact metric space. 
If there is a continuous map from $X$ onto $Y$, then it is mathematically confirmed that $Y$ is homeomorphic to a decomposition space $\mathcal{D}_f = \{ f^{-1}(y) ; y\in Y \}$ of $X$ relative to $f$. 
Hence, we first construct a continuous map on $X$ onto $Y$. 
Since $Y$ is a compact metric space, $Y$ can be covered by the union of finitely many closed sets $Y_1,\dots, Y_{n}$ of $Y$, each diameter of which is less than $1/2$. 
It is mathematically confirmed that there exists a partition $\{X_1,\dots,X_{n}\}$ of $\{0,1\}^\Lambda $ such that 
\begin{equation}
	\left\{
		\begin{array}{lcl}
		X_1=\{0\}_{\lambda _1}\times \{0,1\}^{\Lambda -\{\lambda _1\}}, \\
		X_i=\{1\}_{\lambda _1}\times \cdots \times \{1\}_{\lambda _{i-1}}\times\{0\}_{\lambda _i}\times \{0,1\}^{\Lambda -\{\lambda _1,\dots,	\lambda _i\}}~(i=2,3,\dots,n-1), \\
		X_{n}=\{1\}_{\lambda _1}\times \cdots \times \{1\}_{\lambda _{n-2}}\times\{1\}_{\lambda _{n-1}}\times \{0,1\}^{\Lambda -\{\lambda _1,\dots,\lambda _{n-1}\}},
		\end{array}
	\right.
\label{eqn:2-1}
\end{equation}
where $\lambda _i$ is arbitrarily element of $\Lambda $, $i=1,\dots,n-1$.  
Here, a subset forming $\{k_1\}_{\lambda _1}\times \cdots \times \{k_i\}_{\lambda _{i}}\times \{0,1\}^{\Lambda -\{\lambda _1,\dots,	\lambda _i\}} = \{x : \Lambda \to \{0,1\}, x(\lambda_l)=k_l\in \{ 0,1\}, l=1,\dots,{i}\}$ of a Cantor cube is called a cone. 
Note that each cone is also a 0-dim, perfect, compact Hausdorff-space.
Let $g_1 : X \to \Im(Y)-\{ \o\}$ be defined as following; if $x \in X_i$, then $g_1(x) = Y_i$, for each $i$, where $\Im(Y)$ is the collection of closed sets of $Y$. 
Note that $Y=\cup_{x\in X}g_1(x)$. 
For each $i$, since $Y_i$ is a compact metric space again, there are closed sets $Y_{i_1},\dots,Y_{i_{n_i}}$ of $Y_i$ such that $Y_i = \cup_{j=1}^{n_i} Y_{i_j}$ and each diameter of $Y_{i_j}$ is less than $1/{2^2}$. 
Also, $X_i$ has a partition $\{X_{i_1},\dots,X_{i_{n_i}}\}$ composed of cones such that 
\begin{equation}
	\left\{
		\begin{array}{lcl}
		X_{i_1}=\{1\}_{\lambda _1}\times \cdots \times \{1\}_{\lambda _{i-1}}\times\{0\}_{\lambda _i}\times\{0\}_{\mu _1}\times \{0,1\}^{\Lambda -(\{\lambda _1,\dots,\lambda _i\}\cup \{\mu_1 \})}, \\
		\ \\
		X_{i_j}=\{1\}_{\lambda _1}\times \cdots \times \{1\}_{\lambda _{i-1}}\times\{0\}_{\lambda _i}\times \{1\}_{\mu _1}\times \cdots \times \{1\}_{\mu _{j-1}}\times\{0\}_{\mu _j} \times \{0,1\}^{\Lambda -(\{\lambda _1,\dots,\lambda _i\})\cup (\{\mu _1,\dots,\mu _j\})}~\\
		~~~~~~~~~~~~~~~~~~~~~~~~~~~~~~~~~~~~~~~~~~~~~~~~~~~~~~~~~~~~~~~~~~~~~~~~~~~~~~~~~~~~~~~~~~~~~~~~~~~~~~~~~~(j=2,3,\dots,n_i-1), \\
		X_{i_{n_i}}=\{1\}_{\lambda _1}\times \cdots \times \{1\}_{\lambda _{i-1}}\times\{0\}_{\lambda _i}\times \{1\}_{\mu _1}\times \cdots \times \{1\}_{\mu _{n_i-2}}\times\{1\}_{\mu _{n_i-1}}\times \{0,1\}^{\Lambda -(\{\lambda _1,\dots,\lambda _i\})\cup (\{\mu _1,\dots,\mu _{n_i-1}\})},
		\end{array}
	\right.
\label{eqn:2-2}
\end{equation}
where $\mu _i$ is arbitrarily element of $\Lambda -(\{\lambda _1,\dots,\lambda _i\})$ and each $X_{i_j}$ is 0-dim, perfect, compact Hausdorff-space. 
Let $g_2 : X \to \Im(Y)-\{ \o\}$ be defined as following; if $x \in X_{i_j}$, then $g_2(x) = Y_{i_j}$. 
Then, $Y=\cup_{x\in X}g_2(x)$ and $g_2(x)\subset g_1(x)$ for all $x$. 
Continuing the procedure, we finally obtain a sequence of functions $\{g_n\}$ satisfying that for each $x$ and each $n$, (i) $g_n$ is upper semi-continuous, (ii) $g_{n+1}(x)\subset g_{n}(x)$, (iii) $Y=\cup_{x\in X}g_n(x)$, and (iv) $dia~g_n(x) \to 0$ as $n\to \infty$, where $dia$ stands for diameter of a set. 
Therefore, a continuous map $f$ from $X$ onto $Y$, $x \mapsto \cap_n g_n(x)$ is obtained and the decomposition space $\mathcal{D}_f$ of $X$ relative to $f$ is homeomorphic to $Y$ where $\mathcal{D}_f=\{ f^{-1}(y);y\in Y \}$ with a decomposition topology $\tau(\mathcal{D}_f)=\{\mathcal{U}\subset \mathcal{D}_f; \bigcup \mathcal{U} \in \tau_{0}^{\Lambda} \}$. 
By using the homeomorphism, each point $y$ composing of the geometric structure of $Y$ can be associated with an unique point $f^{-1}(y)$ of the decomposition space $\mathcal{D}_f$ where the relation $f(x)=y$ for $x \in X_{ix}\cap X_{i_{jx}}\cap \cdots$ holds. 

Since geometric structures that we are concerned with in this article are assumed mathematically to be characterized by a finite graph\cite{graph} or a disjoint union\cite{directsum} of points or finite graphs, the construction of a continuous onto map stated above is simplified as the following three steps; 
(i) Divide a Cantor cube into cones such as the relation (\ref{eqn:2-1}) where the number $n$ of cones is, for instance, the number of arcs composing a graph. 
(ii) Construct a decomposition on a cone whose decomposition space is a closed interval $[0,1]$ (or a singleton). 
(iii) Identify the boundaries which consists of end points of arcs with respect to a graphic structure of the geometric pattern.
Note for (ii) that the decomposition space, which is homeomorphic to $[0,1]$, of a cone $\{k_1\}_{\mu _1}\times \cdots \times \{k_i\}_{\mu _{i}}\times \{0,1\}^{\Lambda -\{\mu _1,\dots,	\mu _i\}}$ is obtained from a continuous map $f$ from the cone onto $[0,1]$, the continuous map $f$ being defined by $f(x)=\Sigma _{j=1}^{\infty} a_j/2^j$ such that $x(\lambda _j)=a_j, j=1,2,\dots$ where each $\lambda _j$ is in $\Lambda -\{\mu _1,\dots, \mu _i\}$.
In particular, the decomposition space representing $[0,1]$ of a Cantor cube is obtained practically as the following two cases; putting $M\equiv \{l/2^n ; n=1,2,\dots$ and $ l=1,\dots,2^n-1\}$, (i) for $y=\Sigma _{i=1}^\infty a_i/2^i \not \in M$
\begin{eqnarray}
		f^{-1}(y)=\{a_1\}_{\lambda _1}\times \{a_2\}_{\lambda _2}\times \cdots \times \{0,1\}^{\Lambda -\{a_1,a_2,\cdots\}}, 
		\label{eqn:2-5-1}
\end{eqnarray}
and (ii) for $y = l/2^n\in M$
\begin{eqnarray}
		f^{-1}(y)=\Big{[} \{a_1\}_{\lambda _1}\times \{a_2\}_{\lambda _2}\times \cdots \times \{a_{n-1}\}_{\lambda _{n-1}}\times \{0\}_{\lambda _{n}}\times \{1\}_{\lambda _{n+1}}\times \{1\}_{\lambda _{n+2}}\times  \cdots \times \{0,1\}^{\Lambda -\{\lambda _1,\lambda_2,\cdots\}}\Big{]}
		\nonumber \\
		\cup \Big{[}\{a_1\}_{\lambda _1}\times \{a_2\}_{\lambda _2}\times \cdots \times \{a_{n-1}\}_{\lambda _{n-1}}\times \{1\}_{\lambda _{n}}\times \{0\}_{\lambda _{n+1}}\times \{0\}_{\lambda _{n+2}}\times  \cdots \times \{0,1\}^{\Lambda -\{\lambda _1,\lambda_2,\cdots\}} \Big{]}	
		\label{eqn:2-5-2}
\end{eqnarray}
for some $a_1,\dots, a_{n-1}$. 
Here, $f^{-1}(0)=\{0\}_{\lambda _1}\times \{0\}_{\lambda _2}\times \cdots \times \{0,1\}^{\Lambda -\{\lambda_1,\lambda_2,\cdots\}}$ and $f^{-1}(1)=\{1\}_{\lambda _1}\times \{1\}_{\lambda _2}\times \cdots \times \{0,1\}^{\Lambda -\{\lambda_1,\lambda_2,\cdots\}}$.

Finally, note that the construction process stated in this section is not unique because we can choose the index elements $\lambda _i$'s arbitrarily in $\Lambda$.

\section{Discussion}
\label{sec:3}
In this section, the basic concepts in the previous section are applied to some geometric models with compactness and discuss their geometric structures by using decomposition spaces of $X$. 

\begin{figure}[h]
	\begin{center}
		\includegraphics[clip,width=3cm]{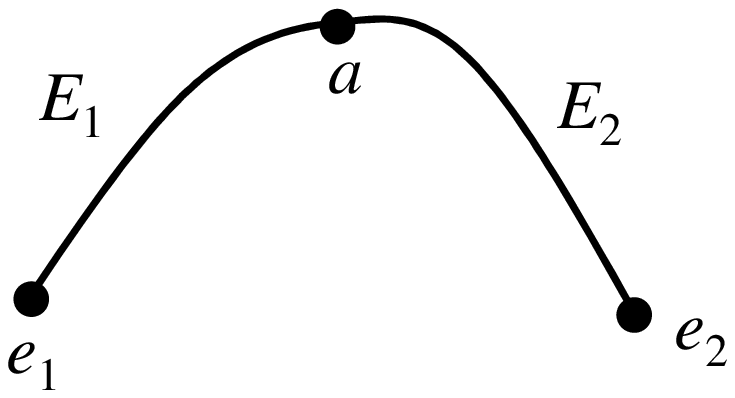}
		\hspace{25mm}
		\includegraphics[clip,width=4cm]{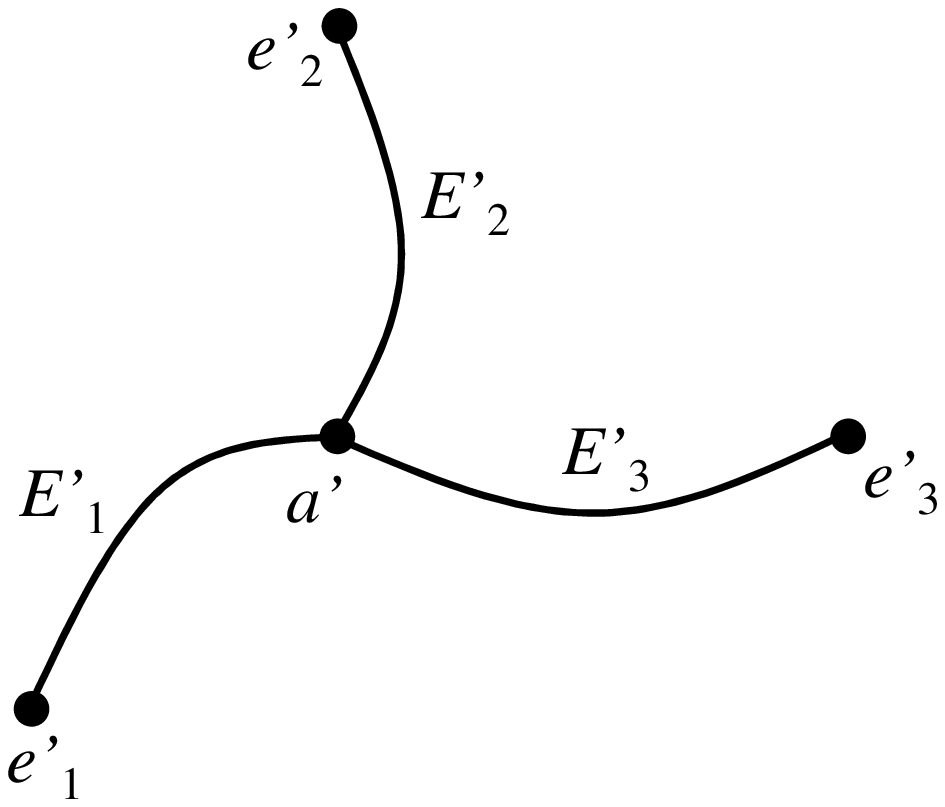}\\
		(a)
		\hspace{69mm}
		(b)
	\end{center}
	\caption{Schematic explanation of two types of network configuration. (a) geometric model $Y_1$; two nodes $e_1$ and $e_2$ are connected by edges $E_1$ and $E_2$ thorough a node $a$. (b) geometric model $Y_2$; three bonds $E_1', E_2'$, and $E_3'$ emanate from a node $a'$ to $e_1', e_2'$, and $e_3'$, respectively. }  
\label{fig:3-1}
\end{figure}

\begin{figure}[h!]
	\begin{center}
		\includegraphics[clip,width=4cm]{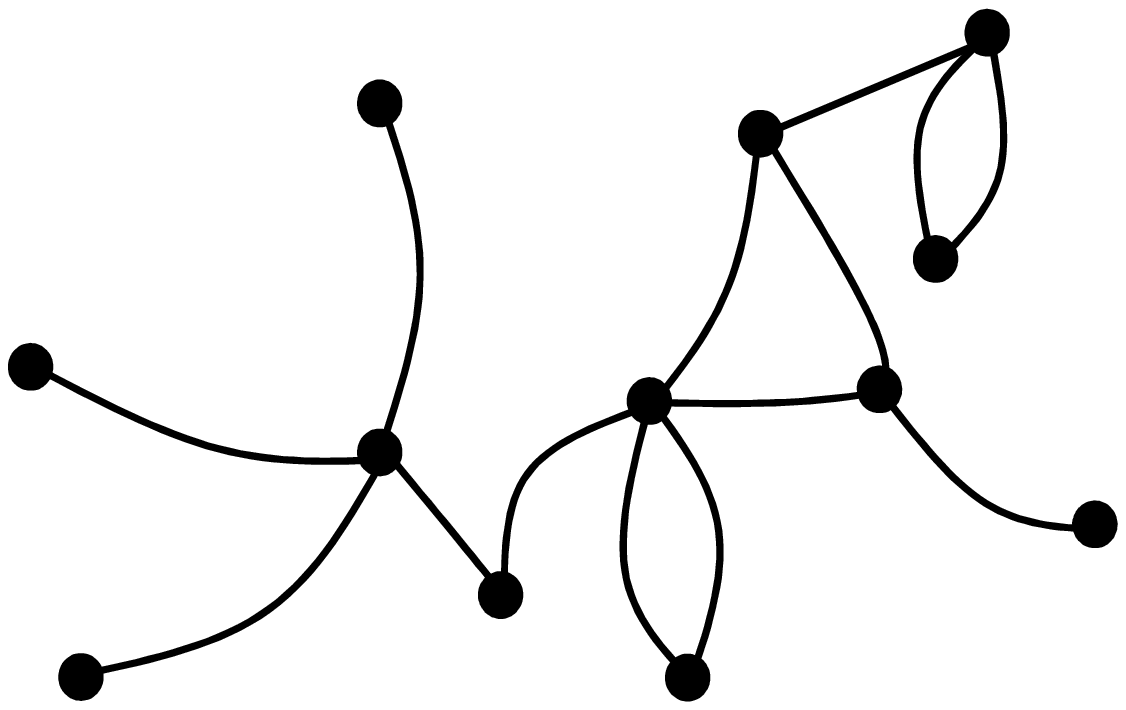}
		\hspace{28mm}
		\includegraphics[clip,width=2.5cm]{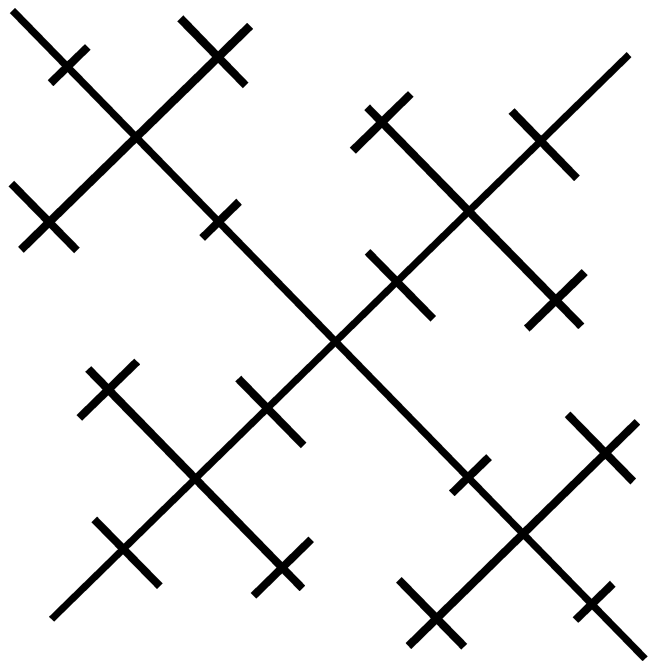}\\
		(a)
		\hspace{60mm}
		(b)
	\end{center}
	\caption{Geometric models of (a) a finite graph $Y_g$ and (b) a dendrite $Y_t$.}  
\label{fig:3-2}
\end{figure}

To begin with, we consider two types $Y_1$ and $Y_2$ with simple network configuration shown in fig.\ref{fig:3-1} ; $Y_1$ is a figure composed of three nodes ${e_1,e_2,a}$ and two bonds $E_1$ and $E_2$ connecting $e_1$ with $a$ and $e_2$ with $a$, respectively. 
$Y_2$ is a figure in which three bonds $E'_1, E'_2,$ and $E'_3$ emanate from a node $a'$ to $e'_1, e'_2$, and $e'_3$, respectively.  
Obviously, $Y_1$ is an arc with end points $e_1$ and $e_2$, whereas, $Y_2$ is regarded as an union of three arcs the intersection of which is just one end point $a'$. 
Since $Y_1$ is an arc, namely, it is homeomorphic to $[0,1]$, the basic concepts obtained the previous section for $Y=[0,1]$ can be directly applied to the space. 
Let $h$ be homeomorphism from $Y_1$ onto $[0,1]$. Then, it is verified that every point $x$ located in $Y_1$ can be written as the point of a decomposition space $\mathcal{D}_1$ of $X=\{0,1\}^\Lambda$ as following two case; (i) if $h(x) \not \in M (\equiv \{l/2^n ; n=1,2,\dots$ and $ l=1,\dots,2^n-1\})$, then 
\begin{eqnarray}
      x \doteq  \{k_1\}_{\lambda _1}\times \{k_2\}_{\lambda _2} \times \cdots \times \{0,1\}^{\Lambda -\{\lambda _1,\lambda _2, \dots \}},
\label{eqn:3-1-1}
\end{eqnarray}
where $k_1, k_2, \dots$ are points in $\{0,1\}$ giving $h(x) = \Sigma _{i=1}^{\infty} k_i/2^i$, and  $\doteq$ is the sign of identification of $x$ with a corresponding point $f^{-1}(x)$ of $\mathcal{D}_1$. 
(ii) if $h(x) \in M$, then
\begin{eqnarray}
      x  \doteq  \Big{[} \{k_1\}_{\lambda _1}\times \{k_2\}_{\lambda _2}\times \cdots \times \{k_m\}_{\lambda _m}\times \{0\}_{\lambda _{m+1}}\times \{1\}_{\lambda _{m+2}}\times \{1\}_{\lambda _{m+3}}\times  \cdots \times \{0,1\}^{\Lambda -\{\lambda _1,\lambda_2,\cdots\}} \Big{]}
			\nonumber \\
		 \cup  \Big{[} \{k_1\}_{\lambda _1}\times \{k_2\}_{\lambda _2}\times \cdots \times \{k_m\}_{\lambda _m}\times \{1\}_{\lambda _{m+1}}\times \{0\}_{\lambda _{m+2}}\times \{0\}_{\lambda _{m+3}}\times  \cdots \times \{0,1\}^{\Lambda -\{\lambda _1,\lambda_2,\cdots\}} \Big{]}
\label{eqn:3-1-2}
\end{eqnarray}
for some $m$, where $k_1, \dots, k_m$ are points in $\{0,1\}$ giving $h(x) \in M$. 
Here, to simplify we introduce a sign $S_x$ defined by 
\begin{equation}
S_x \equiv \left\{
\begin{array}{lcr}
(\ref{eqn:3-1-1}),~~~~ h(x) \not\in M\\
(\ref{eqn:3-1-2}),~~~~ h(x) \in M,
\end{array}
\right.
\label{eqn:3-2}
\end{equation}
and then
\begin{eqnarray}
      x \doteq S_x 
\label{eqn:3-3}
\end{eqnarray}
for $x \in Y_1$.
Note that assuming $h(e_1)=0$ and $h(e_2)=1$, the end points $e_1$ and $e_2$ form
\begin{eqnarray}
      e_1 \doteq \{0\}_{\lambda _1}\times \{0\}_{\lambda _2}\times \cdots \times \{0,1\}^{\Lambda -\{\lambda _1,\lambda_2,\cdots\}}, 
		~e_2  \doteq \{1\}_{\lambda _1}\times \{1\}_{\lambda _2}\times \cdots \times \{0,1\}^{\Lambda -\{\lambda _1,\lambda_2,\cdots\}}.\label{eqn:3-4}
\end{eqnarray}
By the relation (\ref{eqn:3-3}) the geometric feature of $Y_1$ is completely characterized in the decomposition space $\mathcal{D}_1$ of $X$. 
Focusing on $Y_2$, clearly the topology of $Y_2$ differs from $Y_1$ at a point $a'$.   
To characterize the geometric feature of $Y_2$ by applying the basic concepts, we consider a partition $\{X^1, X^2, X^3\}$ of $X$, each cone $X^i$ of which corresponds to one of the three arcs $E'_1\cup \{ e'_1,a' \}, E'_2 \cup \{ e'_2,a' \}, E'_3 \cup \{ e'_3,a' \}$, defined as followings; 
\begin{equation}
	\left\{
		\begin{array}{lcl}
		X^1 = \{0\}_{\mu _1}\times \{0,1\}^{\Lambda -\{\mu_1\}}, \\
		X^2 = \{1\}_{\mu _1}\times \{0\}_{\mu _2}\times \{0,1\}^{\Lambda -\{\mu _1,\mu _2 \}}, \\
		X^3 = \{1\}_{\mu _1}\times \{1\}_{\mu _2}\times \{0,1\}^{\Lambda -\{\mu _1,\mu _2\}},
		\end{array}
	\right.
\label{eqn:3-5}
\end{equation}
where $\mu _i \in \Lambda, i=1,2$. 
Letting $h_i$ be a homeomorphism from $E'_i\cup \{ e'_i,a' \}$ onto $[0,1]$ for $i=1,2,3$ with $h_1(a')=h_2(a')=h_3(a')=0$, it is easily shown from step (ii) and (iii) in previous section that each point $y$ located in $Y_2=\cup_{i=1}^3 [E'_i\cup \{ e'_i,a' \}]$ is represented as followings; for $y \in E'_i\cup \{ e'_i,a' \}$ with $y \not = a'$, then
\begin{eqnarray}
      y \doteq X^{i} \cap S^i_y, ~~(i=1,2,3)
		\label{eqn:3-6}
\end{eqnarray}
where $S^i_y$ is defined by (\ref{eqn:3-2}) for $h_i$ instead of $h$, and 
\begin{eqnarray}
      a' \doteq \cup _{i=1}^3 X^{i} \cap S_{a'},
		\label{eqn:3-7}
\end{eqnarray}
where in this case $S_{a'}=S^1_{a'} =S^2_{a'} =S^3_{a'} = \{0\}_{\lambda _1}\times \{0\}_{\lambda _2}\times \cdots \times \{0,1\}^{\Lambda -\{\lambda _1,\lambda_2,\cdots\}}$ by $h_1(a')=h_2(a')=h_3(a')=0$.
Hence, the relations (\ref{eqn:3-6}) and (\ref{eqn:3-7}) characterizes the geometric feature of $Y_2$ in the decomposition space $\mathcal{D}_2$ of $X$. 
Note in these relations that the elements $\lambda _1, \lambda _2, \dots$ are chosen in $\Lambda - \{\mu _1, \mu _2\}$.
Intuitively, $\cup _{i=1}^3 X^{i}$ in (\ref{eqn:3-7}) represents that the point $a'$ possesses just three arcs emanated from $a'$, and $S_{a'}$ determines the position of $a'$ in each arc. 
From the relations (\ref{eqn:3-6}) and (\ref{eqn:3-7}) for the geometric structure of $Y_2$, a representation of a finite graph shown in Fig. \ref{fig:3-2} (a)  by a decomposition space can be analogized. 
Let us suppose a finite graph $Y_g$ composed of arcs $E_1,\dots, E_r (r<\infty)$. 
To each arc there corresponds a partition $\{ X^1,\dots, X^r \}$ of $X$ such that each $X^i$ is defined as well as that in (\ref{eqn:3-5}) with indexes $\mu_1, \dots, \mu_{r-1} \in \Lambda$. 
Then, it is confirmed that the representations in a decomposition space $\mathcal{D}_g$ for a node $x$ with bonds $E_{t_1},\dots,E_{t_q}$ and a point $y$ in a bond $E_{i}$ are
\begin{eqnarray}
		x \doteq \cup _{j=1}^q (X^{t_j} \cap S^{t_j}_x), y \doteq X^{i} \cap S^{i}_y,
		\label{eqn:3-8}
\end{eqnarray}
respectively. 

\begin{figure}[h!]
	\begin{center}
		\includegraphics[clip,width=4cm]{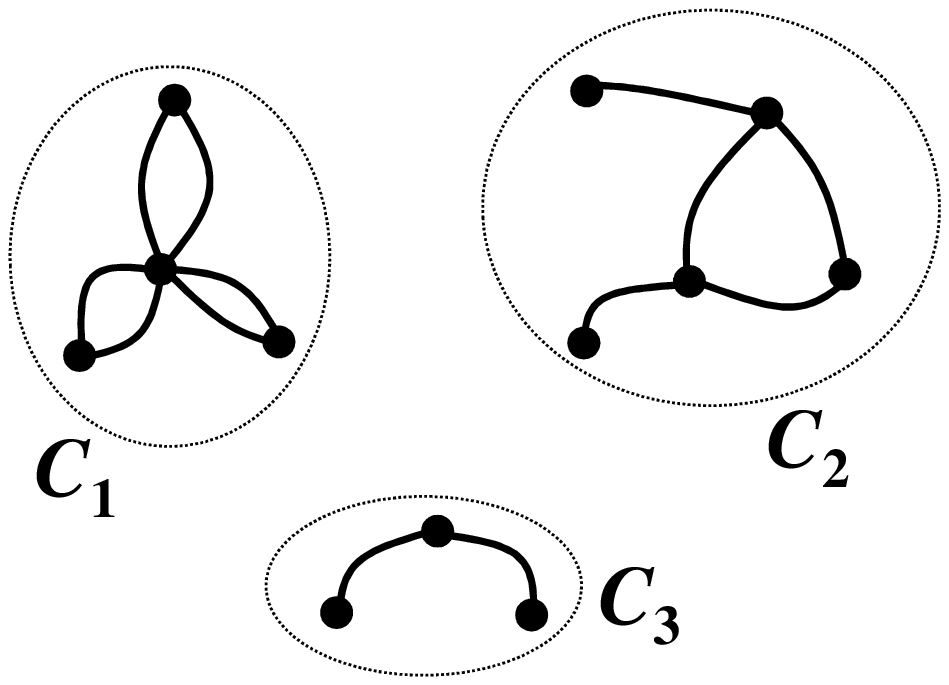}
		\hspace{25mm}
		\includegraphics[clip,width=4cm]{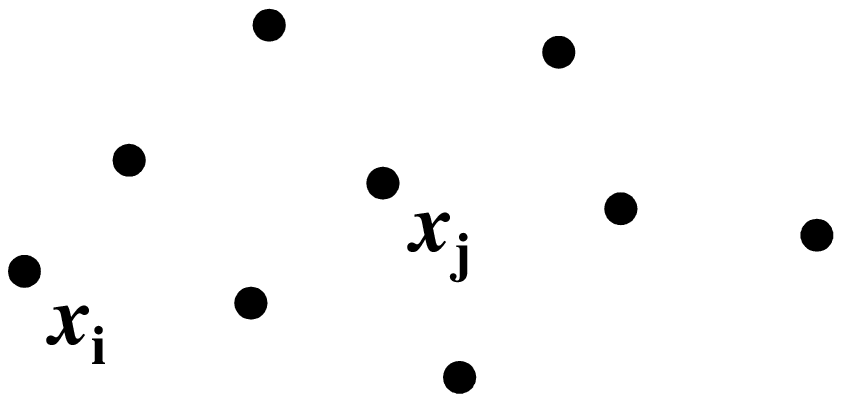}\\
		(a)
		\hspace{60mm}
		(b)
	\end{center}
	\caption{Geometric models of (a) a clusterized structure $Y_c$ where the number of clusters $s=3$, and (b) a totally disconnected clusterized structure $Y_d$.}  
\label{fig:3-3}
\end{figure}

Note that as a example of the graphic structure we can consider a tree such as a dendrite\cite{dendrite}.
A tree is a graph that has no cyclic part shown in (b) of Fig. \ref{fig:3-2}, namely, that does not contains a space homeomorphic to a unit sphere.
In this case, the representation for a tree by a decomposition space $\mathcal{D}_t$ is the same as the relation (\ref{eqn:3-8}). \\ 
~~\\
~~\\
Next, we focus on a geometric model with some clusters, each cluster consisting of a finite graph.
Let $Y_c$ be a topological space described by the geometric figure with clusterized structure Fig. \ref{fig:3-3} (a).
Then, $Y_c$ may have a topological property of a disjoint union of some finite graphs $C_1,\dots, C_s$, namely, we define $Y_c$ to be a topological space $(\bigoplus_{i=1}^sC_i, \bigoplus_{i=1}^s \tau _i)$ where $(\bigoplus_{i=1}^sC_i, \bigoplus_{i=1}^s\tau _i)$ is a disjoint union of a collection of finite graphs $\{(C_i,\tau_{i}),i=1,\dots,s \}$.
Note that since each $C_i$ is a compact metric space, $Y_c=\bigoplus_{i=1}^sC_i$ is a compact metric space. 
Now, let us apply the step (i)-(iii) in Sec 2 to $Y_c$. 
For disjoint clusters $C_1,\dots, C_s$, we first construct a partition $\{J_{1},\dots, J_{s}\}$ of $X$ using new elements ${\xi_1,\dots, \xi_{s-1}}\in \Lambda$ such that
\begin{equation}
	\left\{
		\begin{array}{lcl}
		J_{1} & = & \{0\}_{\xi  _1}\times \{0,1\}^{\Lambda -\{\xi  _1\}}, \\
		J_{j} & = & \{1\}_{\xi  _1}\times \cdots \times \{1\}_{\xi  _{j-1}}\times \{0\}_{\xi  _j}\times \{0,1\}^{\Lambda -\{\xi  _1,\cdots \xi _{j-1}\}} ~~ (j=2,\dots ,s-1), \\
		J_{s} & = & \{1\}_{\xi  _1}\times \cdots \times \{1\}_{\xi  _{s-2}}\times \{1\}_{\xi  _{s-1}}\times \{0,1\}^{\Lambda -\{\xi  _1,\cdots \xi _{s-1}\}},
		\end{array}
	\right.
\label{eqn:3-9}
\end{equation}
each cone $J_i$ corresponding to $C_i$.
As each $C_i$ is a finite graph, by regarding $J_i$ as $X$ in the above discussion about a finite graph, the relation (\ref{eqn:3-8}) is obtained for each $J_i, i=1, \dots, s$.
Therefore, representation of whole space $Y_c$ by a decomposition space $\mathcal{D}_c$ of $X$ is obtained as followings;
assuming that $x\in Y_c$ belongs to a cluster $C_{i_0}$ then
\begin{equation}
		x \doteq J_{i_0} \cap 
		\left\{
		\begin{array}{lcl}
		\cup _{j=1}^q (X^{t_j} \cap S^{t_j}_x), \\
		X^{i} \cap S^i_x,
		\end{array}
	\right.
\label{eqn:3-10}
\end{equation}
where $x$ is a point located either at a node of a finite graph $C_{i_0}$ with bonds $E_{t_1}^{i_0},\dots,E_{t_q}^{i_0}$ or at a point in a bond $E^{i_0}_i$, $E^{i_0}_1, \dots E^{i_0}_{r(i_0)}$ $(t_q\leq r(i_0))$ being arcs composing of a finite graph $C_{i_0}$.  
Note that each $\mu_j$ emerging in each $X^{i}, i=1,\dots, r(j), j = 1,\dots, s$ is in $\Lambda-\{\xi_1,\dots, \xi_s\}$ and each $\lambda _j$ emerging in each $S_x^{l},l=1,\dots, r(j), j = 1,\dots, s$ does not take $\mu_j$ and $\xi _j$ that are already used in $\Lambda$. 
In the relation (\ref{eqn:3-10}), the term $J_{i_0}$ characterizes $x$ belonging to a graphic cluster $C_{i_0}$ and the successive terms characterize a location of $x$ in the graph $C_{i_0}$.

As a special case, we are concerned with a clusterized structure in which each cluster is composed of just one point shown in (b) of Fig.\ref{fig:3-3}.
That is, each graphic structure of $C_i$ is assumed to be an singleton $\{ x_i \}$\cite{note}.
Then, $\bigoplus_{i=1}^sC_i = \cup_{i=1}^s\{x_i\}=\{x_1,\dots,x_s\}$ is a finite totally disconnected compact metric space, denoted by $Y_d$. 
Since each $J_i$ in relation (\ref{eqn:3-10}) means that a given point belongs to $C_i=\{x_i\}$, the decomposition space $\mathcal{D}_d$ for $Y_d$ is obtained by $\mathcal{D}_d = \{J_1,\dots,J_s \}$ having 
\begin{eqnarray}
		x_i \doteq J_i, ~i=1,\dots, s.
		\label{eqn:3-11}
\end{eqnarray}
~~\\

\begin{figure}[h!]
	\begin{center}
		\includegraphics[clip,width=4cm]{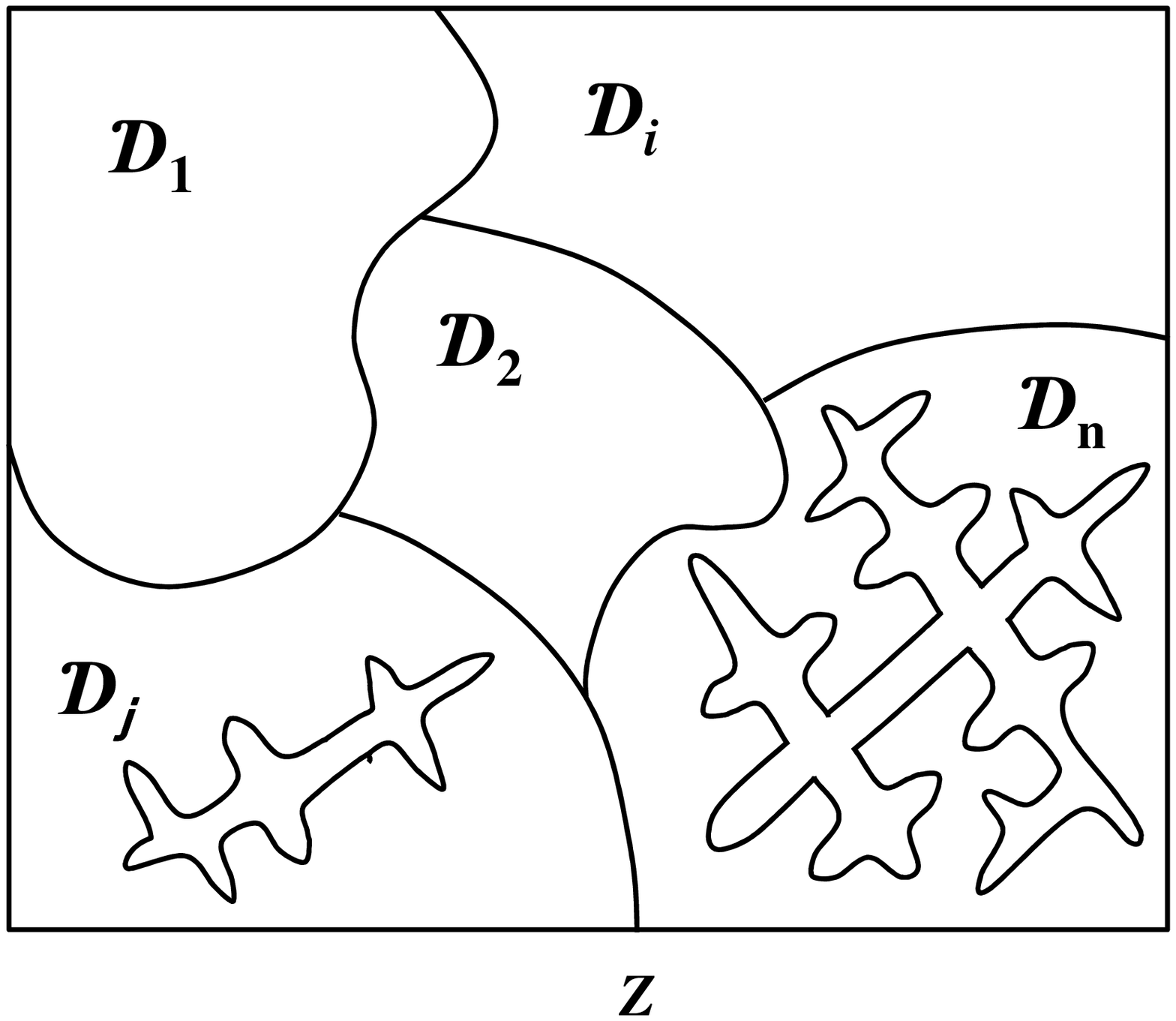}
	\end{center}
	\caption{Schematic explanations of a polycrystal $Z$ filled with dendritic decomposition spaces $\mathcal{D}_i$.}  
\label{fig:3-4}
\end{figure}

Finally, we will show an application of the method of representation by a decomposition space $\mathcal{D}_c$ for a clusterized structure  to the issue stated in Sec. 1 that a polycrystal can be filled with an arbitrary finite number of a crystal characterized by a specific geometric structure, i.e., dendritic, or self-similar structure.
The roughly sketch of situation for the issue is shown in Fig. \ref{fig:3-4}. 
According to the issue we proposed a sufficient condition such that the geometric structure of a polycrystal $Z$ is characterized by a 0-dim, perfect, compact Hausdorff-space such as $(Z,\tau)=(\{0,1\}^{\Lambda},\tau^{\Lambda}_0)$.
If $Z$ satisfies the condition, there exists for arbitrary given number $n$ a partition $\{Z_1,\dots , Z_n\}$ of $Z$ and then each crystal $Z_i$  is 0-dim, perfect, compact Hausdorff-space characterized by its decomposition space $\mathcal{D}_{Z_i}$ with a specific geometric structure.
Since it is mathematically confirmed that $Z$ is filled with the decomposition spaces $\mathcal{D}_{Z_1},\dots,\mathcal{D}_{Z_n}$ in the sense that $Z = \bigcup_{i=1}^{n} \bigcup_{D\in \mathcal{D}_{Z_i}}D$ and $\mathcal{D}_{Z_i}$ and $\mathcal{D}_{Z_j}, i\not = j$ are disjoint, the polycrystal is filled with the crystals each geometric structure of which is characterized by $\mathcal{D}_{Z_i}$. 
In this procedure, it is convinced that the partition $\{Z_1,\dots, Z_n\}$ of $X$ can be taken as the partition $\{J_1,\dots,J_n\}$ defined in (\ref{eqn:3-9}) by $Z_i\equiv J_i$ in assuming the number of clusters is $s=n$.
Suppose that, for instance, dendritic structure characterizes each crystal (Fig.\ref{fig:3-4}).
Then, this situation is rigorously equivalent to that of the clusterized geometric model in which a graphic structure of each cluster is dendritic. 
In other words, we can regard each crystal composing of the polycrystal as a cluster and then the geometric structure of the polycrystal can be described by a kind of clusterized structure. 
By the discussion about the above clusterized structure, we can derive a decomposition space $\mathcal{D}_c$ of $X$ representing the geometric structure of the polycrystal such that each point of $\mathcal{D}_c$ satisfies the relation (\ref{eqn:3-10}) in which the term $J_{i_0}$ shows $x$ to be in a cluster $C_{i_0}$, namely, in a crystal characterized by $\mathcal{D}_{Z_{i_0}}$ in this case. 
Therefore, it follows from the consideration that for each $i=1,\dots, n$,
\begin{eqnarray}
	  \mathcal{D}_{Z_i} = \Big{\{} y \doteq J_i \cap \cup _{j=1}^q (X^{t_j} \cap S^{t_j}_{y}) ; y\in Y_t^i \Big{\}} \cup \Big{\{} y \doteq J_i \cap X^{j} \cap S^j_{y} ; y\in Y_t^i  \Big{\}}. 
		\label{eqn:3-12}
\end{eqnarray}
Also, the following relation is easily obtained;
\begin{eqnarray}
		\mathcal{D}_c = \cup_{i=1}^n \mathcal{D}_{Z_{i}}.
		\label{eqn:3-13}
\end{eqnarray}
Note that it is mathematically confirmed that $Z$ is filled with the decomposition spaces $\mathcal{D}_{Z_i}$ given in (\ref{eqn:3-12}) that are mutually disjoint each other.
The relation (\ref{eqn:3-12}) provides the practical representation of the dendritic crystal $\mathcal{D}_{Z_i}$ which is induced in the discussion to lead the sufficient condition of the issue, and (\ref{eqn:3-13}) provides the relationship of the single crystals to a whole polycrystal composed of them where the geometric structure of the polycrystal is represented by $\mathcal{D}_c$.  
Therefore, the representation of decomposition spaces for the clusterized structure we have shown in this section can be widely applicable to discuss geometric structures of condensed matters from clusterized network configurations to polycystal, noncrystalline or amorphous.\\       
~~\\
In our method discussed here, a Cantor cube is introduced as a conceptional model 
and then we practically obtain an universal representation of geometric structures 
such as the graphic and clusterized structures.
Note that our viewpoint is universally applicable to any condensed matters 
independently of detail internal structures of matters.
Since each character of these geometric structures is connected with a mathematical nature of a Cantor cube, 
new universal properties of condensed matters will be revealed by analyzing a Cantor cube model.
For instance, this approach can contribute to obtaining a mathematical condition 
for determining geometrically configuration of condensed matters, 
as in the case of group theory for mathematical limitation of geometric formation in structural phase transitions of crystals.

\section{Conclusion}
\label{sec:4}

We have shown the method to characterize geometric structures of condensed matters based on a Cantor cube $(X,\tau)=(\{0,1\}^{\Lambda},\tau_0^{\Lambda})$. 
Considering a hierarchic structure of partitions of $X$ composed of cones, any geometric model with compactness can be universally represented as a decomposition space of $X$. 
In this sense, an universal structure exists in disordered geometric formation of condensed matters. 
By using the method, several geometric models such as graphic structures, clusterized structures are represented by corresponding decomposition spaces of $X$. 
In particular, we have also shown a practical form of a decomposition space of polycrystal filled with an arbitrary finite number of  crystal with specific structure i.e., dendritic structure by treating it as a special case of the decomposition representation for the clusterized structural geometric model. 


\begin{thebibliography}{9}
\bibitem{decomposition}
Let $(A,\tau)$ be a topological space. A partition $\mathcal{D}$ of $A$ is a set $\{\o \not = D \in A \}$ of nonempty subsets of $A$ such that $D\cap D' = \o$ for $D \not = D', D,D'\in \mathcal{D}$ and $\bigcup \mathcal{D}(=\cup_{D\in \mathcal{D}})=A$. A decomposition space $(\mathcal{D},\tau(\mathcal{D}))$ of $(A,\tau)$ is a topological space whose topology $\tau(\mathcal{D})$ on a partition $\mathcal{D}$ of $A$ is defined by $\tau(\mathcal{D}) =\{\mathcal{U} \subset \mathcal{D}; \bigcup \mathcal{U} \in \tau \}$. The space $(\mathcal{D},\tau(\mathcal{D}))$ is a kind of quotient space of $(A,\tau)$. See, for the detailed discussions, \cite{Nadler}.
\bibitem{graph}
In a point set topology, a finite graph is defined as a connected compact metric space which is the union of finitely many arcs two of which are either disjoint or intersect only one or both their end points, where a space that is homeomorphic to a closed interval $[0,1]$ is called an arc, and the points corresponding to $0,1$ are called end-points of the arc.
%
Note that each end point in this definition is regarded as a node and an arc eliminating its end points is regarded as a bond in a graph.  
%
See, for the detailed discussions, \cite{Nadler}.
\bibitem{dendrite}
In a point set topology, a dendrite is defined as a topological space which is a Peano continuum (namely, a connected, locally connected, compact, metric-space) containing no simple closed curve. Note that a tree implies a dendrite. See, for the detailed discussions, \cite{Nadler}.
\bibitem{note}
Strictly speaking, a singleton is not a graph. However, since the geometric model $Y_d$ is definable as a finite metric space, it can be regarded as a special case of a model with clusterized structure.
\end{thebibliography}
\end{document}